\documentclass[preprint2]{aastex}

\slugcomment{To appear in ApJ Letter}

\shorttitle{Solar flares and global oscillations in the Sun}
\shortauthors{Karoff \& Kjeldsen}

\begin{document}

\title{Evidence that solar flares drive global oscillations in the Sun}

\author{C. Karoff \& H. Kjeldsen}
\affil{Department of Physics and Astronomy, University of Aarhus, DK 8000 Aarhus C, Denmark}

\begin{abstract}
Solar flares are large explosions on the SunÕs surface caused by a sudden release of magnetic energy. They are known to cause local short-lived oscillations travelling away from the explosion like water rings. Here we show that the energy in the solar acoustic spectrum is correlated with flares. This means that the flares drive global oscillations in the Sun in the same way that the entire Earth is set ringing for several weeks after a major earthquake like the December 2004 Sumatra-Andaman Earthquake. The correlation between flares and energy in the acoustic spectrum of disk-integrated sunlight is stronger for high-frequency waves than for ordinary $p$-modes which are excited by the turbulence in the near surface convection zone immediately beneath the photosphere. \end{abstract}

\keywords{Sun: flares --- Sun: helioseismology}

\section{Introduction}
The global 5-minute $p$-modes oscillations of the Sun have in the latest three decades provided us with valuable information about the physics inside the Sun. The oscillations are believed to be excited by the turbulence in the near surface convection zone immediately beneath the photosphere \citep{1996Sci...272.1296G}. Complete reflection is expected to take place only for oscillations with frequencies lower than the solar atmospheric acoustic cut-off (about 5.3 mHz). Oscillations with higher frequencies, which are seen in the Sun up to 9 mHz (sometimes known as high-frequency waves) are only partially reflected and have transmitted components, which propagate upwards through the atmosphere \citep{1990ApJ...362..256B}. While models of near surface convection can now explain the excitation and damping of the ordinary $p$-modes no consensus has been reached about the details of what drives oscillations with frequencies larger than the acoustic cut-off frequency in the solar atmosphere. Different studies have proposed different models of the high-frequency waves \citep{1990ApJ...362..256B, 1991ApJ...375L..35K, 1996ApJ...456..399J}, but none has explained all the features seen in the observations.

We have used data from the SOHO (Solar and Heliospheric Observatory) \citep{1995SoPh..162..101F} and the GOES (Geostationary Operational Environmental Satellite) \citep{1994SoPh..154..275G}  satellites to show that there is a strong correlation between the energy at high frequency in the solar acoustic spectrum and the appearance of solar flares. The correlation suggests that the driving of the high frequency waves is related to solar flares. 
Such a driving mechanism is well known from earthquakes where the entire Earth is set ringing for several weeks after a major earthquake like the December 2004 Sumatra-Andaman Earthquake \citep{2005Sci...308.1139P}. In fact it was shown already in 1972 that solar flares could excite free global oscillations in the Sun in the same way large earthquakes excite free global oscillation in the Earth \citep{1972ApJ...176..833W}. Since then a few attempts have been made in order to find a correlation between the energy of the low degree $p$-modes and flares but none of them have been very successful \citep{1999MNRAS.303L..63G, 2004SoPh..220..307C, 2006SoPh..238..219A}. Instead flares are found to cause local short-lived high-degree oscillations in the active regions in which they are produced \citep{1998Natur.393..317K}. 

\begin{figure}[t]
\centering
\epsscale{1.0}
\plotone{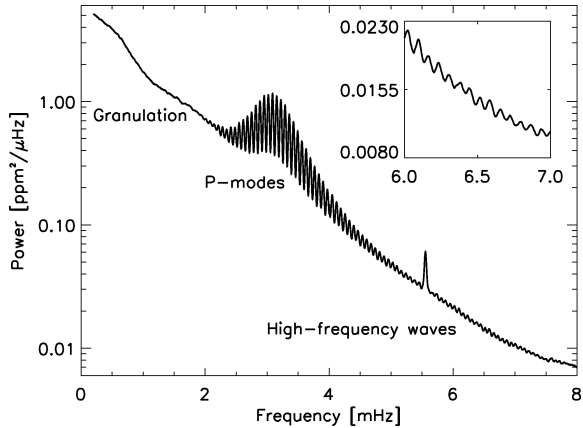}
\caption{Smoothed power density spectrum of the Sun. The spectrum consists of granulation noise (below 1.8 mHz), acoustic power from completely trapped $p$-modes (between 1.8 and 5.3 mHz), and power from the high-frequency waves (with frequency higher than 5.3 mHz). The peak at 5.55 mHz is an artifact. }
\label{fig2}
\end{figure}

High-frequency waves were discovered in high-degree observations of the Sun \citep{1988ApJ...334..510L}, but recently they have also been seen in disk integrated data such as the BISON radial velocity data \citep{2003ESASP.517..247C}, the GOLF radial velocity data \citep{2005ApJ...623.1215J}  and the VIRGO intensity data \citep{2005ApJ...623.1215J}.  As the high-frequency waves have the highest S/N in intensity \citep{2005ApJ...623.1215J} we have chosen to focus on Virgo data (green channel) in this study. Fig.~1 shows a power density spectrum of the solar data. Here the high-frequency waves appear clearly from 5.3 mHz up to almost 8 mHz. 

\begin{figure*}[t]
\centering
\epsscale{2.0}
\plotone{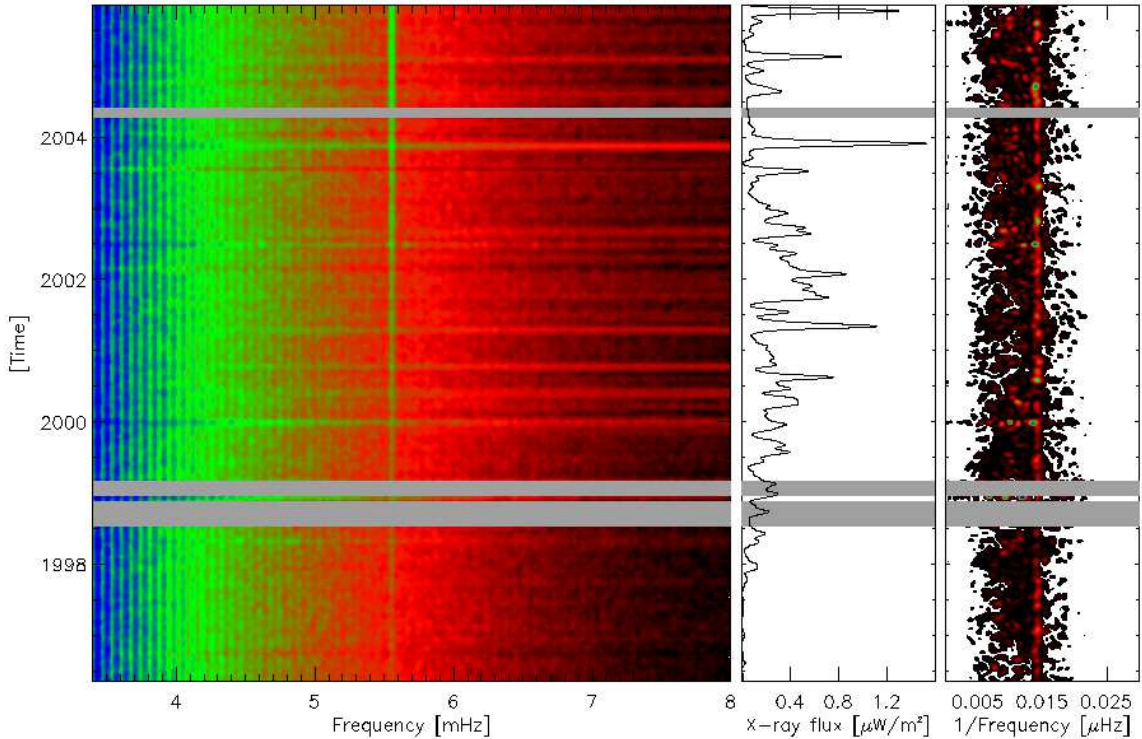}
\caption{Smoothed frequency/time diagram of the Sun from disc-integrated data from the VIRGO instrument on SOHO is shown in the left panel. The clear vertical lines in the left of the image are the $p$-mode oscillations, and it is evident that these lines continue far beyond 5.3 mHz. The vertical line at 5.55 mHz is an instrumental artifact. The bright horizontal lines are strongly correlated with solar flares, represented by the X-ray flux in the middle panel. These lines are most prominent in the high-frequency region of the spectrum. The right panel shows the amplitude spectra of the individual power density spectra in the left panel, with a bright line at the frequency separation of the high-frequency waves of 0.014 $\mu$Hz$^{-1}$ (70 $\mu$Hz). The gray regions mark times with no data from SOHO. The color scale is logarithmic.}
\label{fig1}
\end{figure*}

\section{Frequency/time diagrams}
We have calculated frequency/time diagrams of the integrated sunlight in order to evaluate the temporal behaviour of the high-frequency waves. This has been achieved by calculating power spectra of a number of running substrings of a long time series from the VIRGO instrument on SOHO. The power spectra were calculated as least squares spectra \citep{1976Ap&SS..39..447L} in order to insure proper treatment of the gapes in the data. We tested different length of the substrings and found 7.5 days to be the smallest length that we could use and still have enough S/N in the spectra to see the high-frequency waves clearly. In order to keep the rest of the analysis as simple as possible we calculated the spectra with a frequency resolution of 1 $\mu$Hz. The power spectra were converted to power density spectra by multiplying a given power spectrum with the effective length of a given substring, which we calculated as the reciprocal of the area under the spectral window (in power) of the substring \citep{2005ApJ...635.1281K}. 

Successive, overlapping, substrings were displaced by 18 hours with respect to each other. The procedure yielded 4620 power spectra. The frequency/time diagrams can now be constructed by stacking the power spectra vertically, producing a power diagram with frequency varying along the horizontal axis and time on the vertical axis. In order to enhance the visibility the diagrams were smoothed by a Gaussian point-spread-function having widths of 33 $\mu$Hz in frequency and 24 days in time. The widths were chosen to be as small as possible and still allow the structures of the high-frequency waves to appear as clear as possible.

As the analyzed VIRGO data have a sampling of 60 sec and thus a Nyquist frequency of 8.3 mHz we have chosen to calculate the power spectra up to 8 mHz. There are some gaps in the data set, mainly around the SOHO vacation in the summer of 1998, but when we smooth the diagram interpolation is made of these  gaps by the Gaussian point-spread-function. As we generally do not think that this interpolation is reliable we have remove the data in the gaps from Figs.~2 \& 3.

The frequency/time diagrams in Fig.~2 exhibit a regular pattern of vertical bright lines at low frequency. These indicate the power in the ordinary $p$-modes. But the pattern continues beyond 5.3 mHz, demonstrating that underlying mode structure is present also in the high-frequency waves.
The frequency/time diagrams also show a pattern of horizontal lines. It is evident from the middle panel that this pattern is correlated with solar flares. The number and strength of the flares has been represented by the X-ray flux measured in the soft channel (1--8 \AA ) of the X-ray sensors on the GOES satellites \citep{1994SoPh..154..275G}. The X-ray data have been smoothed with a Gaussian running mean of width 25 days in accord with the resolution of the oscillation data. The width was taken to be 25 rather than 24 days as the Virgo data to some extend has been smoothed two times, first by calculating a power spectrum of a substring of length 7.5 days and second by smoothing with a Gaussian point-spread-function of width 24 days in the vertical direction. 

\begin{figure}[t]
\centering
\epsscale{1.0}
\plotone{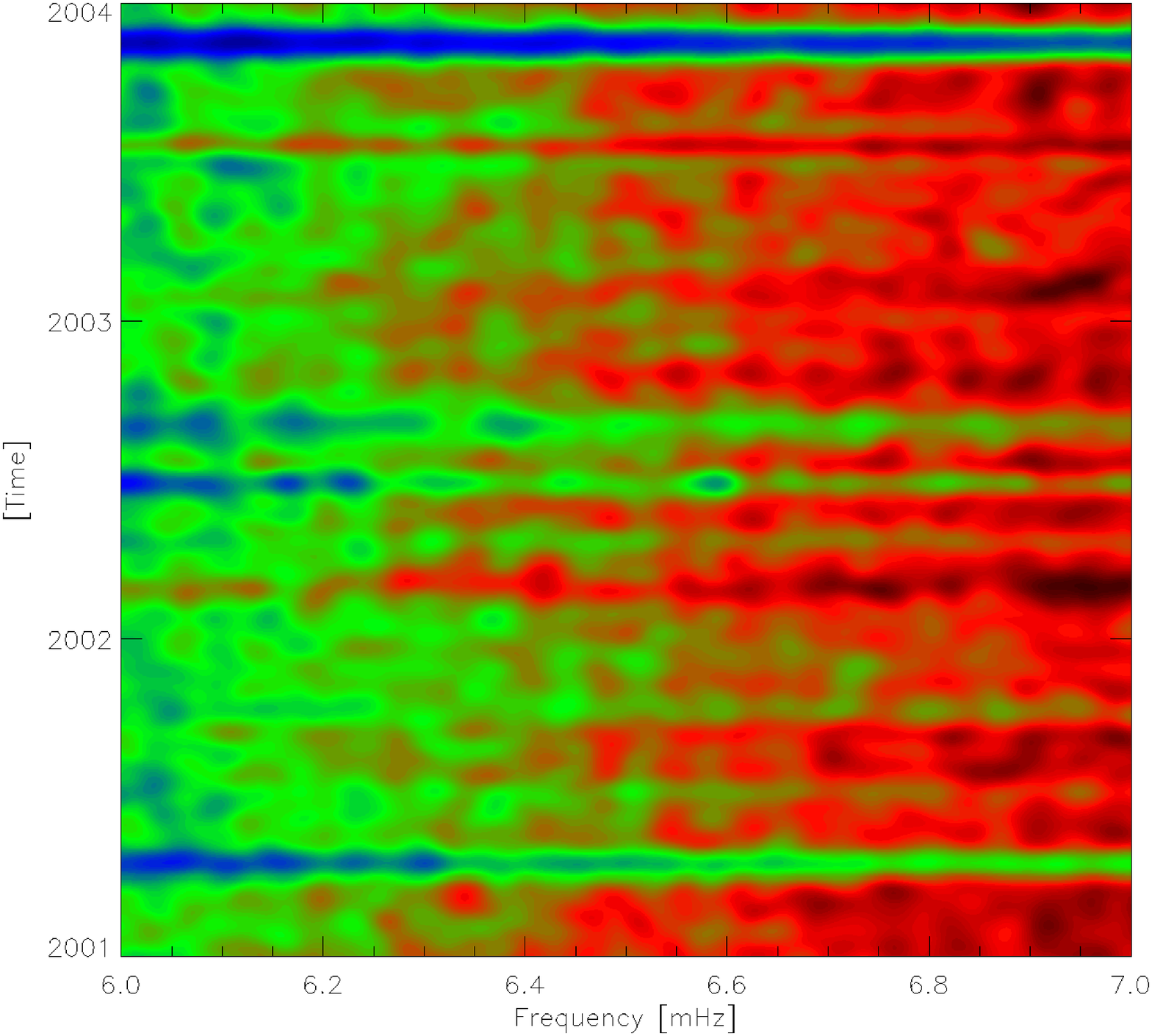}
\caption{A zoom in on Fig.~2 where the regularity of the high-frequency waves appears more clearly.}
\label{fig2}
\end{figure}

\begin{figure}[t]
\centering
\epsscale{1.0}
\plotone{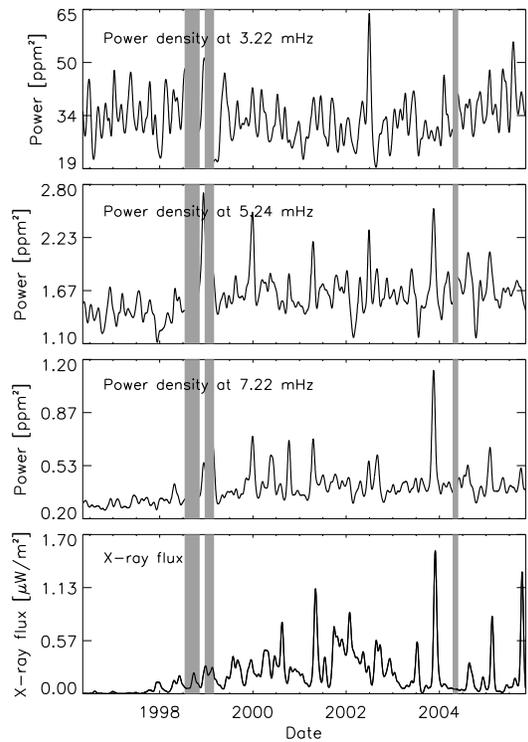}
\caption{The X-ray flux from the Sun (lower panel), total power of the mode at 7.22 mHz (second  panel), at 5.24 mHz (third panel) and at 3.22 mHz (top panel). It is evident that the correlation between the total power of the modes and the X-ray flux increases with increasing frequency. The slow change of the total power of the mode at 3.22 mHz is a result for the solar dynamo \citep[][and references herein]{2003ApJ...582L.115C}.} 
\label{fig2}
\end{figure}

\section{Correlation between solar flares and power density at different frequencies}
In order to analyze how flares correlate with oscillations at different frequencies we have compared the variation of the total power within 20 $\mu$Hz of three different modes (at 3.22, 5.24 and 7.22 mHz). It is seen in Fig.~3 that the correlation is strongest for the high-frequency oscillations at 7.22 mHz, at 5.24 the correlation is much weaker and at 3.22 the dominant feature is the well known slow variation of the dynamo \citep[][and references herein]{2003ApJ...582L.115C}. All the three modes have higher power density around 2006 than around 1996, though the power is expected to be the same as both times are close to solar minimum. This trend is also seen in the X-ray flux (the mean power density is higher between 2004 and 2006 than between 1996 and 1998), but as the trend is monochromic increasing it could be an instrumental effect. Though we do not know of such an effect in the instrument. 

In Fig.~4 we show the general correlation between the power density and X-ray flux independent of the individual modes. In order not to have the correlation dominated by the long period trends we have subtracted a Gaussian running mean with a width of 2 years from both the power density and the X-ray flux.  Fig.~4 shows that the temporal behavior of the power density spectrum is more correlated with the temporal behavior of X-ray flares at high-frequency than at low frequency.

We do not know the exact excitation mechanism for $p$-modes being excited by flares, however two possible effects may be considered in order to explain and understand the behavior of the high-frequency $p$-mode power. First of all the observed power of $p$-modes at high-frequency are significantly lower than the power per mode observed at the peak (3 mHz). According to \citet{2001ApJ...546..585S} one find the $p$-mode power to decrease inversely to the fourth power of the frequency (above 3 mHz) simply as a consequence of the granulation power decreasing as $\nu^{-4}$. Therefore a single event (like a flare) that excites a mode will have a larger relative effect at high frequencies because the other excitation sources are much smaller. Related to this one should also note that the high-frequency background (between $p$-modes) show an increased correlation with flare activity. One may therefore argue that the background noise from granulation correlate with flare activity (as clearly seen in Figs.~2 and 5) and the increased background transfer more power into the high-frequency modes via the stochastic excitation, which is exactly what is observed. At frequencies around 3 mHz one do not find the background to correlate as strongly (see Fig.~5) as at higher frequencies (5-7 mHz).

This may explains why attempts to see a correlation between the energy of the ordinary $p$-modes and flares have not been very successful \citep{1999MNRAS.303L..63G, 2004SoPh..220..307C, 2006SoPh..238..219A}. Fig.~4 shows a steady clime of correlation over the acoustic cut-off frequency, not a sudden jump. This is in agreement with what we expect as the reflectivity of the solar atmosphere is expected to decrease continuously over the acoustic cut-off frequency. 

\begin{figure}[t]
\centering
\epsscale{1.0}
\plotone{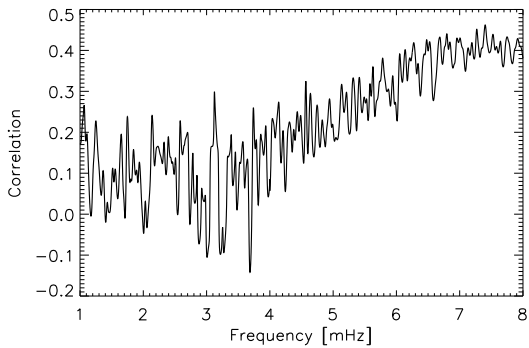}
\caption{Correlation between the power density and the X-ray flux from the Sun as a function of frequency.}
\label{fig3}
\end{figure}

\section{Correlation between solar flares and global oscillations at high frequency}
The acoustic spectrum contains two parts at high frequency: a background and the acoustic waves. In order to separate these two parts we have preformed the same analysis as preformed by \citet{2005ApJ...623.1215J} and calculated amplitude spectra of the individual power density spectra in the frequency/time diagram. The exponential decay of the power density spectra were removed before these amplitude spectra were calculated as done by \citet{2005ApJ...623.1215J}. The amplitude spectra as function of time are shown in the right panel of Fig.~2. It is seen that the frequency separation of the high-frequency waves remains constant at 70 $\mu$Hz throughout the observation independent of solar flares. 

\section{Discussion}
The correlation we see between the power density of high-frequency wave and the flares is similar to what one would expect from a comet impact. This case was analyzed for the Shoemaker-Levy impart on Jupiter by \citet{1994MNRAS.269L..17G}. In that paper it was also noted that such impacts on the Sun would have the largest response in the high-frequency domain. 
The discovery presented here will improve our understanding of the flares. An advance here will be provided by the asteroseismic data that we shall receive from the Kepler satellite \citep{2003SPIE.4854..129B}. High-frequency waves have been detected in asteroseismic data of the solar-like stars beta Hydri and alpha Cen A \& B \citep{2007MNRAS.381.1001K}, and with the Kepler data we should be able to observe the time variation of the amplitudes of such waves in a large number of solar-like stars, perhaps suggesting the presence of flares in those stars. The last outcome could aid our understanding of stellar dynamos.

\acknowledgments
The authors would like thank D.O. Gough, J. Christensen-Dalsgaard, T. Bedding and T. Arentoft for comments. The authors acknowledge use of computing resources from the Danish Center for Scientific Computing. CK acknowledges support from the Danish AsteroSeismology Centre and the Instrument Center for Danish Astrophysics. SOHO is a mission of international cooperation between the European Space Agency and NASA. The National Oceanic and Atmospheric Administration operates GOES.

\clearpage

\end{document}